\documentclass[preprint,showpacs,amsmath,amssymb,preprintnumbers]{revtex4}
\usepackage{mathrsfs}

\usepackage{graphicx,color}
\usepackage{dcolumn}
\usepackage{bm}
\usepackage{CJK}

\begin{document}
\preprint{RIKEN-iTHEMS-Report-17}
\preprint{RIKEN-QHP-347}

\title{Nuclear mass predictions based on Bayesian neural network approach with pairing and shell effects}

\author{Z. M. Niu$^{1,2}$}\email{zmniu@ahu.edu.cn}
\author{H. Z. Liang$^{2,3,4}$}\email{haozhao.liang@riken.jp}

\affiliation{$^1$School of Physics and Materials Science, Anhui University,
             Hefei 230601, China}
\affiliation{$^2$Interdisciplinary Theoretical Science Research Group,
             RIKEN, Wako 351-0198, Japan}
\affiliation{$^3$RIKEN Nishina Center,
             Wako 351-0198, Japan}
\affiliation{$^4$Department of Physics, Graduate School of Science,
             The University of Tokyo, Tokyo 113-0033, Japan}

\date{\today}

\begin{abstract}
Bayesian neural network (BNN) approach is employed to improve the nuclear mass predictions of various models.
It is found that the noise error in the likelihood function plays an important role in the predictive performance of the BNN approach.
By including a distribution for the noise error, an appropriate value can be found automatically in the sampling process, which optimizes the nuclear mass predictions.
Furthermore, two quantities related to nuclear pairing and shell effects are added to the input layer in addition to the proton and mass numbers.
As a result, the theoretical accuracies are significantly improved not only for nuclear masses but also for single-nucleon separation energies.
Due to the inclusion of the shell effect, in the unknown region, the BNN approach predicts a similar shell-correction structure to that in the known region, e.g., the predictions of underestimation of nuclear mass around the magic numbers in the relativistic mean-field model.
This manifests that better predictive performance can be achieved if more physical features are included in the BNN approach.
\end{abstract}

\maketitle


Mass is a fundamental property of atomic nuclei. It can be employed to extract various nuclear structure information, such as nuclear pairing correlation, shell effect, deformation transition, and so on~\cite{Lunney2003RMP}. Nowadays it has been also widely used to determine nuclear effective interactions~\cite{Bender2003RMP}. Moreover, nuclear mass is essential to determine the nuclear reaction energy in astrophysics and hence plays a crucial role in understanding the origin of elements in Universe~\cite{Burbidge1957RMP}. In addition, the accurate mass determination is very important to test the unitarity of Cabibbo-Kobayashi-Maskawa matrix~\cite{Liang2009PRC, Hardy2015PRC}.

Measurements of nuclear mass have achieved great progress in recent years~\cite{Franzke2008MSR, Sun2015FP} and about $3000$ nuclear masses have been measured up to now~\cite{Wang2017CPC}. However, the accurate predictions of nuclear mass are still a great challenge for theoretical models, due to the difficulties in the exact theory of nuclear interaction and in the quantum many-body calculations. Nowadays three types of nuclear models are mainly used in global mass predictions: macroscopic, macroscopic-microscopic, and microscopic mass models. The Bethe-Weizs\"{a}cker (BW) mass formula is the first model used to estimate nuclear masses~\cite{Weizsacker1935ZP, Bethe1937RMP}, which belongs to the macroscopic type. It assumes the nucleus is similar to a charged liquid drop, so the microscopic effects, such as shell effect, cannot be well described.
By taking into account the important corrections related to the microscopic effects, the macroscopic-microscopic models are developed, such as the finite-range droplet model (FRDM)~\cite{Moller2012PRL} and the Weizs\"{a}cker-Skyrme (WS) model~\cite{Wang2014PLB}. The microscopic mass models are mainly rooted in the density functional theory, which are more complicated but potentially have a better ability of extrapolation. In the non-relativistic framework, a series of Hartree-Fock-Bogoliubov (HFB) mass models have been constructed with the Skyrme~\cite{Goriely2009PRLSkyrme, Goriely2016PRC} or Gogny~\cite{Goriely2009PRLGogny} effective interactions. In recent years, the relativistic mean-field (RMF) model also receives wide attention due to its success in describing various nuclear phenomena~\cite{Vretenar2005PRp, Meng2006PPNP, Meng2016Book, Meng2006PRC, Liang2008PRL, Niu2013PRCR, Niu2017PRC} and its successful applications in astrophysics~\cite{Sun2008PRC, Niu2009PRC, Xu2013PRC, Niu2013PLB}. Based on the RMF model, global calculations of nuclear mass have been made and the accuracies were gradually improved~\cite{Geng2005PTP, Hua2012SCPMA, Arteaga2016EPJA}.

The accuracies of these mass models range from about $3$~MeV for the BW model~\cite{Kirson2008NPA} to about $0.3$~MeV for the WS model~\cite{Wang2014PLB}. However, these accuracies are still insufficient to the studies of exotic nuclear structures and astrophysics nucleosynthesis. Especially, these models predict very different nuclear masses with the differences even up to tens of MeV when they are extrapolated to the neutron drip line. Therefore, it is still a high demand to further improve the existing nuclear mass models. Some techniques have been developed along this direction, such as the radial basis function (RBF) approach~\cite{Wang2011PRC, Niu2013PRCb, Zheng2014PRC, Niu2016PRC} and the image reconstruction technique with the CLEAN algorithm~\cite{Morales2010PRC}. Moreover, the neural network has been proved to be a very powerful tool and it has been widely used in an impressive range of problem domains, such as pattern recognition and machine learning, see, e.g., Books~\cite{Haykin2009Book, Bishop2006Book} and the references therein. The application of neural network to predict nuclear masses can be traced back to the 1990s~\cite{Gazula1992NPA}. A series of works after that were developed to further improve its predictive performance~\cite{Gernoth1993PLB, Athanassopoulos2004NPA, Zhang2017JPG}. It was also extended to study other nuclear properties, such as nuclear $\beta$-decay half-lives~\cite{Costiris2009PRC}. These approaches usually need many parameters, in general hundreds or even thousands of parameters, for achieving better predictions, so the over-fitting problem and the quantification of uncertainties in the predictions should be treated in a reliable way.

The Bayesian approach can avoid the over-fitting problem by introducing the prior distribution of parameters, and it can quantify the uncertainties in the predictions since all parameters have probability distributions~\cite{Neal1996Book}. Thus, it would be a valuable approach for improving the mass predictions of nuclear models. However, the Bayesian approach involves high-dimensional integrals over the whole parameter space, so its calculations are very time-consuming and great progress was achieved only in the last decades along with the developments in sampling methods and dramatic improvements in the speed and memory of computers~\cite{Bishop2006Book}. Recently, the Bayesian neural network (BNN) approach was applied to improve the theoretical predictions of nuclear masses~\cite{Utama2016PRC} and nuclear charge radii~\cite{Utama2016JPG}. The noise error in the likelihood function is a key quantity in the BNN approach, however, it was usually much simplified by taking a fixed value in the previous studies~\cite{Utama2016PRC, Utama2016JPG}. In this work, we will introduce a prior distribution for the noise error. Furthermore, only the proton and mass numbers were considered in the input layer of the neural network in the previous studies~\cite{Utama2016PRC, Utama2016JPG}. Here we will consider more physical features into the input layer, i.e., we will include two quantities related to the well known nuclear pairing and shell effects, and investigate their influences on the predictive performance of the BNN approach.

In the Bayesian approach, the model parameters $\bm{\omega}$ are described probabilistically. A probability distribution $p(\bm{\omega})$ is introduced over all possible values of $\bm{\omega}$ based on our background knowledge, which is called the \textit{prior} distribution. When we observe a set of data $D=\{(\bm{x_1}, t_1), (\bm{x_2}, t_2), ..., (\bm{x_N}, t_N)\}$, this distribution will be updated by using the Bayes' theorem
\begin{eqnarray}
  p(\bm{\omega}|D) = \frac{p(D|\bm{\omega})p(\bm{\omega})}{p(D)}\propto p(D|\bm{\omega})p(\bm{\omega}),
\end{eqnarray}
where $\bm{x_n}$ and $t_n$ $(n=1, 2, ..., N)$ are input and output data, $N$ is the number of data; $p(D|\bm{\omega})$ is the likelihood function, which contains the information about parameters $\bm{\omega}$ derived from the observations; $p(\bm{\omega}|D)$ is the probability distribution of parameters $\bm{\omega}$ after the data $D$ are considered, which is called the \textit{posterior} distribution; $p(D)$ is a normalization constant, which ensures the posterior distribution is a valid probability density and integrates to one.

For the likelihood function $p(D|\bm{\omega})$, a Gaussian distribution, $p(D|\bm{\omega}) = \exp(-\chi^2/2)$, is usually employed, where the objective function $\chi^2$ reads
\begin{eqnarray}\label{Eq:ObjFun}
  \chi^2=\sum_{n=1}^N \left(\frac{t_n - S(\bm{x}; \bm{\omega})}{\Delta t_n}\right)^2.
\end{eqnarray}
Here, the standard deviation parameter $\Delta t_n$ is the associated noise error related to the $n$th observable. For the BNN approach, the function $S(\bm{x}; \bm{\omega})$ is described with a neural network, which is
\begin{eqnarray}\label{Eq:NeuralNetwork}
  S(\bm{x}; \bm{\omega})=a+\sum_{j=1}^H b_j \tanh\left(c_j+\sum_{i=1}^I d_{ji} x_i\right),
\end{eqnarray}
where $\bm{x}=\{x_i\}$ and $\bm{\omega}=\{a, b_j, c_j, d_{ji}\}$, and $H$ and $I$ are the numbers of neurons in the hidden layer and the number of input variables, respectively.
In total, the number of parameters in this neural network is $1+(2+I)*H$.

For the prior distributions $p(\bm{\omega})$ of model parameters, they are usually set as Gaussian distributions with zero means. However, the precisions (inverse of variances) of these Gaussian distributions are not set as fixed values by hand. We set them as gamma distributions so that the precisions can vary over a large range and hence the BNN approach can search the optimal values of precisions in the sampling process automatically.

After specifying the likelihood function $p(D|\bm{\omega})$ and the prior distribution $p(\bm{\omega})$, the posterior distribution $p(\bm{\omega}|D)$ of model parameters is known in principle. One can then make predictions based on this posterior distribution,
\begin{eqnarray}\label{Eq:AverInBNN}
  \langle S \rangle = \int S(\bm{x}; \bm{\omega})p(\bm{\omega}|D) d\bm{\omega}.
\end{eqnarray}
Since the model parameters are described with a probability distribution, an estimate of uncertainty in theoretical predictions is obtained naturally as
\begin{eqnarray}
  \Delta S = \sqrt{\langle S^2 \rangle - \langle S \rangle^2 }.
\end{eqnarray}
Note that Eq.~\eqref{Eq:AverInBNN} involves a high-dimensional integral in the whole parameter space. For that, we will employ the Monte Carlo integral algorithm, where the posterior distribution $p(\bm{\omega}|D)$ is sampled using the flexible Bayesian model developed by Neal~\cite{Neal1996Book}, in which the Markov chain Monte Carlo algorithm is employed.

In this work, we will employ the BNN approach to reconstruct mass residuals between experimental data $M^{\exp}$ and mass predictions $M^{\rm th}$ of various models, i.e.,
\begin{eqnarray}
  t_n = M^{\exp}(\bm{x}) - M^{\rm th}(\bm{x}).
\end{eqnarray}
As in Refs.~\cite{Utama2016PRC, Utama2016JPG}, the inputs are usually taken as $\bm{x}=(Z, A)$. However, we will consider more physical information into the BNN approach, so two extra inputs $\delta$ and $P$ related to nuclear pairing and shell effects are also included, which are
\begin{eqnarray}\label{Eq:InputsDeltaP}
  \delta=[(-1)^Z+(-1)^N]/2, \quad P=\nu_p\nu_n/(\nu_p+\nu_n).
\end{eqnarray}
Here, $\nu_p$ and $\nu_n$ are the differences between the actual nucleon numbers $Z$ and $N$ and the nearest magic numbers ($8$, $20$, $28$, $50$, $82$, $126$ for protons and $8$, $20$, $28$, $50$, $82$, $126$, $184$ for neutrons)~\cite{Kirson2008NPA}. For simplicity, we will use BNN-I2 and BNN-I4 to denote the BNN approaches with $\bm{x}=(Z, A)$ and $\bm{x}=(Z, A, \delta, P)$, respectively. Their numbers of neurons are taken as $H=42$ and $H=28$, respectively, so the model parameters in both neural networks are the same as $169$.

The experimental masses are taken from the atomic mass evaluation of 2016 (AME2016)~\cite{Wang2017CPC}, while only those nuclei with $Z, N\geqslant 8$ and experimental errors $\sigma^{\exp}\leqslant 100$~keV are considered. There are $2272$ data left that compose the entire data set. In order to examine the validity of the BNN approach, we separate the entire set into two different sets: the learning set and the validation set. The learning set is built by randomly selecting $1800$ nuclei from the entire set and the remaining $472$ nuclei compose the validation set. For the theoretical mass models, we take two microscopic (RMF~\cite{Geng2005PTP} and HFB-31~\cite{Goriely2016PRC}), two macroscopic-microscopic (WS4~\cite{Wang2014PLB} and FRDM12~\cite{Moller2012PRL}), and two macroscopic (BW~\cite{Kirson2008NPA} and BW2~\cite{Kirson2008NPA}) mass models as examples.


The noise errors in Eq.~\eqref{Eq:ObjFun} were usually taken as a fixed value estimated from mass differences between experimental data and model predictions~\cite{Utama2016PRC, Utama2016JPG}. A more elegant way is to set it as a distribution, and the sampling process can search an appropriate value automatically, which can optimize the nuclear mass predictions. In this work, we will use a gamma distribution for the noise precision (inverse of squared noise error $1/\Delta t^2$), because the gamma distribution is the conjugate prior distribution of the precision of Gaussian distribution, which can make calculations easier in mathematics~\cite{Bishop2006Book}.

\begin{table}
\begin{center}
\caption{
Root-mean-square (rms) deviations (in MeV) of nuclear mass with respect to the experimental data in the learning sets for various mass models and their counterparts improved by the BNN-I2 approach.
For each model, the original rms deviation is denoted by $\sigma_{\rm pre}$, the corresponding posterior rms deviations with a constant and a gamma distribution for noise precision are denoted by $\sigma_{\rm post}^{\rm fixed}$ and $\sigma_{\rm post}^{\rm gamma}$, respectively. The last column shows the reductions from $\sigma_{\rm post}^{\rm fixed}$ to $\sigma_{\rm post}^{\rm gamma}$.} \label{Tab:NoisePrec}
\begin{tabular}{lccccc}
\hline \hline
Model   &~~     &$\sigma_{\rm pre}$   &$\sigma_{\rm post}^{\rm fixed}$    &$\sigma_{\rm post}^{\rm gamma}$    &$\Delta\sigma (\%)$ \\
\hline
RMF     &~~     &~~2.269~~     &~~0.732~~     &~~0.443~~     &~~39.5~~ \\
HFB-31  &~~     &~~0.560~~     &~~0.354~~     &~~0.296~~     &~~16.5~~ \\
WS4     &~~     &~~0.286~~     &~~0.203~~     &~~0.178~~     &~~12.5~~ \\
FRDM12  &~~     &~~0.570~~     &~~0.282~~     &~~0.208~~     &~~26.3~~ \\
BW      &~~     &~~3.236~~     &~~1.035~~     &~~0.850~~     &~~17.9~~ \\
BW2     &~~     &~~1.627~~     &~~0.497~~     &~~0.313~~     &~~37.0~~ \\
\hline \hline
\end{tabular}
\end{center}
\end{table}

Table~\ref{Tab:NoisePrec} gives the root-mean-square (rms) deviations of nuclear mass with respect to the experimental data in the learning sets for various mass models and their counterparts improved by the BNN-I2 approaches. Clearly, the BNN approach can significantly improve the mass predictions even with a fixed noise precision. By using a gamma distribution, the improvements are further enhanced and the reduction in the rms deviations even approaches $40\%$ for the RMF and BW2 models. In the following, all calculations will be performed with a gamma distribution for the noise precision.

\begin{table}
\begin{center}
\caption{Rms deviations (in MeV) of nuclear mass with respect to the experimental data in the learning and validation sets for various mass models and their counterparts improved by the BNN-I2 and BNN-I4 approaches.
The last two columns show the reductions of the rms deviations.} \label{Tab:MassBNN}
\begin{tabular}{lcccccc}
\hline \hline
        &~~
             &$\sigma_{\rm pre}$
             &$\sigma_{\rm post}^{I=2}$
             &$\sigma_{\rm post}^{I=4}$
             &$\Delta\sigma^{I=2}(\%)$
             &$\Delta\sigma^{I=4}(\%)$ \\
\hline
        &~~    &\multicolumn{5}{c}{Learning set} \\
RMF     &~~  &~~2.269~~     &~~0.443~~     &~~0.367~~     &~~80.5~~     &~~83.8~~ \\
HFB-31  &~~  &~~0.560~~     &~~0.296~~     &~~0.246~~     &~~47.1~~     &~~56.1~~ \\
WS4     &~~  &~~0.286~~     &~~0.178~~     &~~0.176~~     &~~37.7~~     &~~38.3~~ \\
FRDM12  &~~  &~~0.570~~     &~~0.208~~     &~~0.187~~     &~~63.5~~     &~~67.1~~ \\
BW      &~~  &~~3.236~~     &~~0.850~~     &~~0.266~~     &~~73.7~~     &~~91.8~~ \\
BW2     &~~  &~~1.627~~     &~~0.313~~     &~~0.247~~     &~~80.8~~     &~~84.8~~ \\
\hline
        &~~    &\multicolumn{5}{c}{Validation set} \\
RMF     &~~  &~~2.242~~     &~~0.480~~     &~~0.415~~     &~~78.6~~     &~~81.5~~ \\
HFB-31  &~~  &~~0.559~~     &~~0.363~~     &~~0.316~~     &~~35.1~~     &~~43.4~~ \\
WS4     &~~  &~~0.283~~     &~~0.222~~     &~~0.212~~     &~~21.8~~     &~~25.3~~ \\
FRDM12  &~~  &~~0.599~~     &~~0.268~~     &~~0.252~~     &~~55.2~~     &~~58.0~~ \\
BW      &~~  &~~3.048~~     &~~0.924~~     &~~0.310~~     &~~69.7~~     &~~89.8~~ \\
BW2     &~~  &~~1.690~~     &~~0.369~~     &~~0.284~~     &~~78.1~~     &~~83.2~~ \\
\hline \hline
\end{tabular}
\end{center}
\end{table}

It is well known that nuclear pairing and shell effects play very important roles in mass predictions~\cite{Lunney2003RMP}. For further improving the mass deviations related to such effects, two extra inputs $\delta$ and $P$ are included in addition to $Z$ and $A$. The corresponding rms deviations for various mass models improved by the BNN-I4 approach are given in Table~\ref{Tab:MassBNN}. The results of original models and those improved by BNN-I2 are also shown for comparison.

As the best example, the liquid-drop BW mass model only includes the volume, surface, symmetry, and Coulomb terms, while both pairing and shell effects are fully neglected~\cite{Kirson2008NPA}. Improved by BNN-I2, its posterior rms deviation is still much larger than those of other mass models. However, with the BNN-I4 approach, its posterior rms deviation is significantly reduced from $850$ to $266$~keV.

\begin{figure}
\includegraphics[width=6cm]{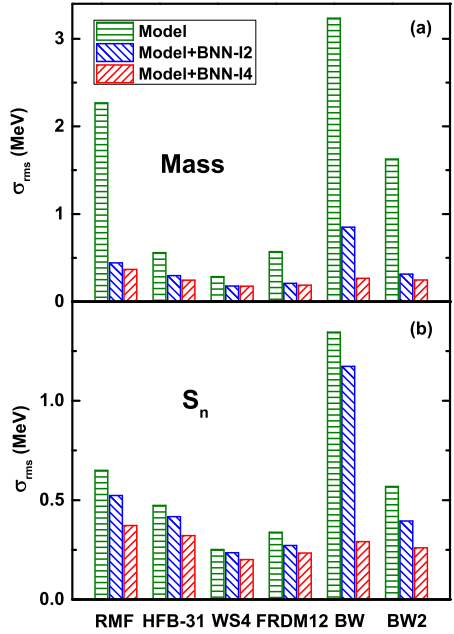}\\
\caption{(Color online) Rms deviations of (a) nuclear mass and (b) single-neutron separation energy with respect to the experimental data in the learning set for various mass models and those improved by the BNN approaches.}\label{Fig1:rmsMassSnLinear}
\end{figure}

In general, improved by the BNN-I4 approach, the rms deviations of all mass models are significantly reduced, e.g., exceeding $90\%$ for the BW model. It can be seen clearly in Fig.~\ref{Fig1:rmsMassSnLinear}(a). In addition, from the rms deviations for the validation set shown in Table~\ref{Tab:MassBNN}, one can evaluate the predictive performance of the BNN approach. Although the rms deviations for the validation set are slightly larger than those for the learning set, the improvements on the original models are still significant.

\begin{table*}
\begin{center}
\caption{Rms deviations (in MeV) of single-neutron ($S_n$), single-proton ($S_p$), two-neutron ($S_{2n}$), and two-proton ($S_{2p}$) separation energies with respect to the experimental data in the learning and validation sets for various mass models and their counterparts improved by the BNN-I2 and BNN-I4 approaches.} \label{Tab:SnSpS2nS2pBNN}
\begin{tabular}{lccccccccccccccc}
\hline \hline
        &~~          &\multicolumn{4}{c}{Model}                   &~~       &\multicolumn{4}{c}{Model+BNN-I2}            &~~        &\multicolumn{4}{c}{Model+BNN-I4}             \\
\cline{3-6}\cline{8-11}\cline{13-16}
        &~~  &$S_n$       &$S_p$       &$S_{2n}$    &$S_{2p}$     &~~    &$S_n$       &$S_p$       &$S_{2n}$    &$S_{2p}$    &~~    &$S_n$       &$S_p$       &$S_{2n}$    &$S_{2p}$   \\
\hline
        &~~                                                               &\multicolumn{14}{c}{Learning set} \\
RMF     &~~  &~0.650~     &~0.818~     &~0.847~     &~1.090~      &~~    &~0.523~     &~0.622~     &~0.420~     &~0.442~     &~~    &~0.372~     &~0.395~     &~0.398~     &~0.472~    \\
HFB-31  &~~  &~0.473~     &~0.507~     &~0.469~     &~0.510~      &~~    &~0.417~     &~0.391~     &~0.349~     &~0.318~     &~~    &~0.322~     &~0.319~     &~0.326~     &~0.298~    \\
WS4     &~~  &~0.252~     &~0.258~     &~0.266~     &~0.306~      &~~    &~0.235~     &~0.216~     &~0.205~     &~0.194~     &~~    &~0.201~     &~0.208~     &~0.215~     &~0.223~    \\
FRDM12  &~~  &~0.339~     &~0.347~     &~0.438~     &~0.430~      &~~    &~0.272~     &~0.266~     &~0.244~     &~0.230~     &~~    &~0.234~     &~0.229~     &~0.252~     &~0.245~    \\
BW      &~~  &~1.345~     &~1.439~     &~1.223~     &~1.245~      &~~    &~1.174~     &~1.237~     &~0.396~     &~0.364~     &~~    &~0.291~     &~0.290~     &~0.355~     &~0.364~    \\
BW2     &~~  &~0.569~     &~0.619~     &~0.858~     &~0.929~      &~~    &~0.395~     &~0.401~     &~0.346~     &~0.337~     &~~    &~0.260~     &~0.264~     &~0.313~     &~0.320~    \\
\hline
        &~~                                                               &\multicolumn{14}{c}{Validation set} \\
RMF     &~~  &~0.626~     &~0.768~     &~0.868~     &~0.977~      &~~    &~0.479~     &~0.571~     &~0.376~     &~0.406~     &~~    &~0.363~     &~0.388~     &~0.418~     &~0.429~    \\
HFB-31  &~~  &~0.390~     &~0.354~     &~0.439~     &~0.390~      &~~    &~0.370~     &~0.329~     &~0.359~     &~0.375~     &~~    &~0.357~     &~0.259~     &~0.349~     &~0.371~    \\
WS4     &~~  &~0.247~     &~0.254~     &~0.256~     &~0.309~      &~~    &~0.247~     &~0.239~     &~0.227~     &~0.257~     &~~    &~0.254~     &~0.238~     &~0.215~     &~0.303~    \\
FRDM12  &~~  &~0.324~     &~0.310~     &~0.459~     &~0.337~      &~~    &~0.246~     &~0.246~     &~0.280~     &~0.248~     &~~    &~0.257~     &~0.258~     &~0.303~     &~0.264~    \\
BW      &~~  &~1.262~     &~1.292~     &~1.135~     &~1.148~      &~~    &~1.136~     &~1.204~     &~0.430~     &~0.353~     &~~    &~0.263~     &~0.336~     &~0.358~     &~0.399~    \\
BW2     &~~  &~0.517~     &~0.589~     &~0.722~     &~0.896~      &~~    &~0.370~     &~0.343~     &~0.284~     &~0.402~     &~~    &~0.304~     &~0.220~     &~0.288~     &~0.351~    \\
\hline \hline
\end{tabular}
\end{center}
\end{table*}

The single-nucleon separation energies are related to the derivatives of nuclear mass surface. They are also very important to nucleon-capture reactions in astrophysics.
Therefore, it is interesting to investigate the improvements of single-nucleon separation energies with the BNN approaches. Previous studies found that the RBF approach is one of the powerful techniques to improve the mass predictions of nuclear models~\cite{Wang2011PRC, Niu2013PRCb, Zheng2014PRC}, but its improvement in overall mass predictions even deteriorates the description of single-nucleon separation energy ($S_n$ or $S_p$) unless the RBF is done twice separately~\cite{Niu2016PRC}. Table~\ref{Tab:SnSpS2nS2pBNN} shows the rms deviations of $S_n$ and $S_p$ with respect to the data in the learning and validation sets for various mass models and their counterparts improved by the BNN approaches. For completeness, the two-neutron ($S_{2n}$) and two-proton ($S_{2p}$) separation energies are given together. The results for the learning set are shown in Fig.~\ref{Fig1:rmsMassSnLinear}(b). It is clear that the BNN approach can improve the predictions of nuclear masses and the single-nucleon separation energies simultaneously, remarkably for the BNN-I4 approach. This indicates the BNN-I4 approach is more effective to simultaneously improve the descriptions of nuclear mass surface and its derivatives than the BNN-I2 approach.

\begin{figure*}
\includegraphics[width=16cm]{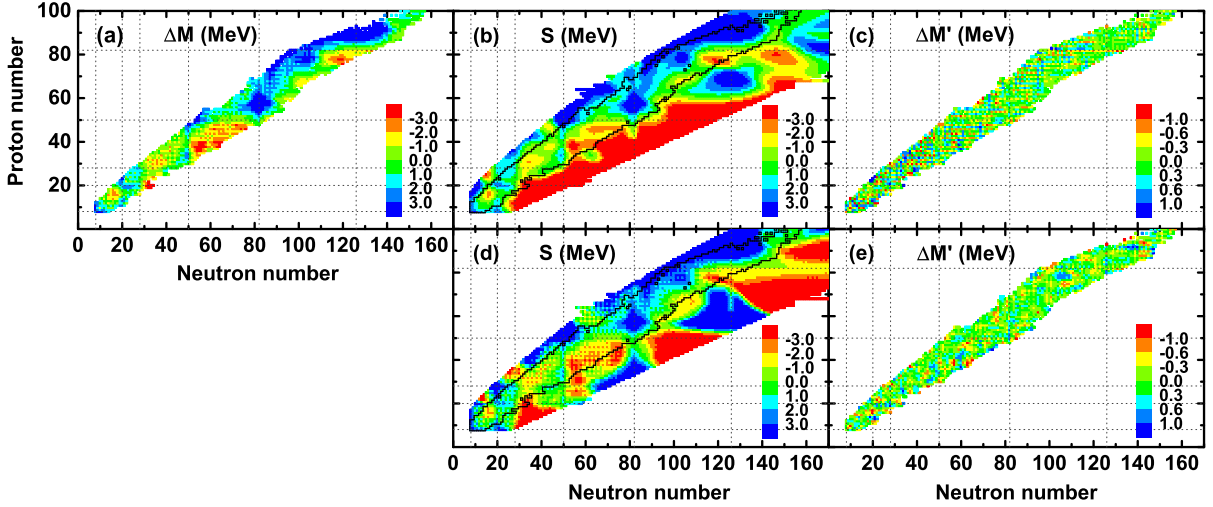}
\caption{(Color online) (a) Mass differences between the experimental data and the RMF predictions in the entire set. (b) Mass corrections $S(Z, A)$ of RMF by the BNN-I2 approach, where the boundary of nuclei in the entire set are shown by the black contours. (c) Mass differences between the experimental data and the RMF predictions improved by the BNN-I2 approach. Panels (d) and (e) are the same as panels (b) and (c), but for the BNN-I4 approach.}\label{Fig2:DiffMRMFBNN}
\end{figure*}

The rms deviation provides only a gross assessment of the accuracy of a nuclear mass model. To show some details, we present the mass differences between the experimental data and the predictions of each nucleus in the entire set in panel (a) of Fig.~\ref{Fig2:DiffMRMFBNN} by taking the RMF mass model as an example. Clearly, there are some large differences, such as in the region around the magic numbers. These discrepancies around the magic numbers were also found in the HFB mass models with Skyrme force~\cite{Kortelainen2010PRC} or Gogny force~\cite{Goriely2016EPJA}, which are generally explained as being due to the physics missing from the energy density functionals---the so-called ``beyond mean-field'' physics. The idea of the BNN approach is to employ a neural network for simulating such kinds of missing physics in nuclear mass models, so it is expected that the mass predictions of nuclear models can be improved. Panel (b) gives the mass corrections $S(Z, A)$ of RMF by using the BNN-I2 approach. It is found that there are very similar structures between panel (a) and those inside the contour lines of panel (b). This indicates the BNN approach can well describe the smooth deviations between the experimental data and theoretical predictions. The mass differences between the experimental data and the mass predictions improved by the BNN-I2 approach are shown in panel (c). Clearly, the mass deviations of the RMF model are almost eliminated. Quantitatively, the resulting rms deviation is reduced from $2.263$ to $0.451$~MeV. However, the remaining differences still show some odd-even staggering structures, i.e., smaller and larger differences appear alternately. In addition, from the structure outside the contour lines in panel (b), the BNN approach predicts a systematic overestimation (underestimation) of nuclear mass in the neutron-rich (neutron-deficient) region except for heavy neutron-rich nuclei. It is different from the structure in the known region inside the contour lines, which holds richer features and predicts an overestimation of nuclear masses for nuclei around the magic numbers.

It is well known the odd-even staggering and local structures around magic numbers are related to nuclear pairing correlation and shell effect, respectively. Therefore, the inclusion of $\delta$ and $P$ in Eq.~\eqref{Eq:InputsDeltaP} is expected to work out these problems. The corresponding results for the BNN-I4 approach are shown in panes (d) and (e) of Fig.~\ref{Fig2:DiffMRMFBNN}. From panel (d), it is clear that the BNN-I4 approach eliminates not only the smooth deviations but also the odd-even staggering. Therefore, there is no remarkable odd-even staggering for the mass differences left in panel (e). Furthermore, the BNN corrections outside the contour lines in panel (d) show more structure features than those in panel (b). For example, it predicts an overestimation of mass for nuclei towards $(Z,N)=(28, 82)$ and $(50, 126)$. This may indicate the extrapolation ability of BNN-I4 is more reliable than that of BNN-I2.

\begin{figure}
\includegraphics[width=6cm]{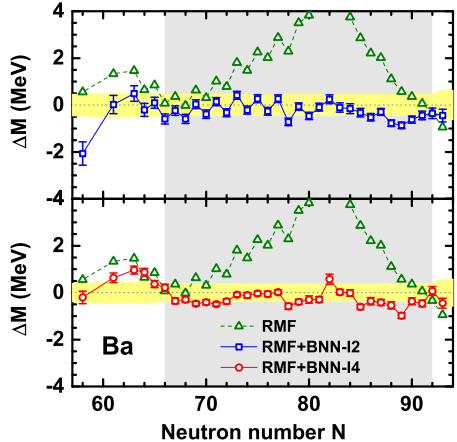}
\caption{(Color online) Mass differences between the experimental data and the RMF predictions without and with the BNN approaches for the Ba isotopes. The range of the entire set is shown in the gray-hatched region. The yellow-hatched region gives the mass uncertainties from the average errors of theoretical models (rms deviations of theoretical models) and the experimental errors. The rms deviations used in panels (a) and (b) are those improved by the BNN-I2 and BNN-I4 approaches, respectively.}\label{Fig3:BaRMF}
\end{figure}

\begin{figure}
\includegraphics[width=6cm]{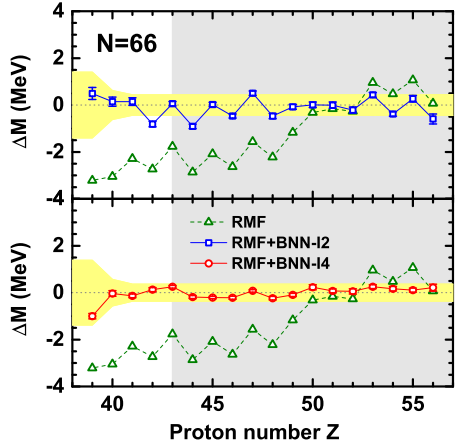}
\caption{(Color online) Same as Fig.~\ref{Fig3:BaRMF}, but for the $N=66$ isotones.}\label{Fig4:Neq66RMF}
\end{figure}

To evaluate the extrapolation ability of BNN approach, we will use those nuclei in AME2016 but not selected into the entire set, since their present experimental errors $\sigma^{\exp}> 100$~keV. Taking the Ba isotopes and $N=66$ isotones as examples, the corresponding results are shown in Figs.~\ref{Fig3:BaRMF} and \ref{Fig4:Neq66RMF}, respectively.
The gray-hatched regions denote the range of the entire set. Clearly, both BNN-I2 and BNN-I4 approaches can eliminate the smooth mass deviations to a large extent, while the BNN-I4 approach remarkably reduces the odd-even staggering. If the extrapolation is not far away from the training region, i.e., when the change in neutron or proton number is not larger than $5$, the RMF mass predictions are well improved by the BNN approaches, especially by BNN-I4. This further manifests the BNN-I4 approach achieves better predictive performance than the BNN-I2 approach.

Apart from improving the mass predictions of nuclear models, the BNN approach also provides the uncertainties in mass predictions, which are shown in Figs.~\ref{Fig3:BaRMF} and \ref{Fig4:Neq66RMF} as well. It is found that the uncertainties of BNN mass predictions become larger and larger for both BNN approaches if they are extrapolated away from training region, while the uncertainties in the BNN-I4 approach are smaller than those in the BNN-I2 approach. In addition, the yellow-hatched regions in Figs.~\ref{Fig3:BaRMF} and \ref{Fig4:Neq66RMF} give the mass uncertainties from the average errors of theoretical models and the experimental errors. We found that the BNN mass predictions generally agree well with the data within uncertainties, even they are extrapolated from the training region. This demonstrates the BNN approaches can estimate uncertainties in mass predictions in quite a reliable way. However, if it is extrapolated too far away from the known region, there might be some new physics effects, which are hidden in the known region and hence cannot be discovered by training the neural network using the known data.


In summary, we have employed the Bayesian neural network approach to improve the nuclear mass predictions of various models. By using a distribution for the noise error in likelihood function, the BNN approach can find the optimal value of the noise error automatically, which improve nuclear mass predictions remarkably.
To better describe nuclear pairing and shell effects on mass predictions, we further include two relevant quantities in addition to the proton and mass numbers, keeping the number of parameters unchanged. It is found that the present BNN approach not only eliminates the smooth mass deviations significantly but also remarkably reduces the odd-even staggering in mass deviations. As a result, the accuracies of all mass models considered here are significantly improved not only for the nuclear masses but also for the separation energies.
Furthermore, the mass corrections with the present BNN approach show more structure features, e.g., it predicts an overestimation of nuclear masses for nuclei towards $(Z,N) = (28,82)$ and $(50,126)$ in the RMF mass model. This manifests better predictive performance can be achieved not only in the known region but also in the unknown region far from the $\beta$-stability line, if more physical features are included in the BNN approach.

It is known that there exists an exact universal energy density functional for nuclear ground-state properties, though it is very difficult even impossible to construct it.
If one is able to find an accurate energy density functional with the BNN approach by taking various densities as the direct inputs, one can make reliable predictions for various nuclear ground-state properties. Works along this line are now in progress. In addition, one can also apply the BNN approach to improve other nuclear properties with many experimental data, such as nuclear charge radii, $\beta$-decay half-lives, and so on.

\section*{Acknowledgements}
We are grateful to Professor T. Hatsuda and Dr. Y. F. Niu for the fruitful discussions. This work was partly supported by the National Natural Science Foundation of China under Grant No.~11205004, the Natural Science Foundation of Anhui Province under Grant No.~1708085QA10, the Key Research Foundation of Education Ministry of Anhui Province under Grant No.~KJ2016A026, and the RIKEN iTHES project and iTHEMS program.





\begin{thebibliography}{99}
\bibitem{Lunney2003RMP}D. Lunney, J. M. Pearson, C. Thibault, Rev. Mod. Phys. \textbf{75}, 1021 (2003).

\bibitem{Bender2003RMP}M. Bender and P.H. Heenen, and P.G. Reinhard, Rev. Mod. Phys. \textbf{75}, 121 (2003).

\bibitem{Burbidge1957RMP}E. Margaret Burbidge, G. R. Burbidge, William A. Fowler, and F. Hoyle, Rev. Mod. Phys. \textbf{29}, 547 (1957).

\bibitem{Liang2009PRC}H. Z. Liang, N. Van Giai, J. Meng, Phys. Rev. C \textbf{79}, 064316 (2009).

\bibitem{Hardy2015PRC}J. C. Hardy, I. S. Towner, Phys. Rev. C \textbf{91}, 025501 (2015).

\bibitem{Franzke2008MSR}B. Franzke, H. Geissel, G. M\"{u}nzenberg, Mass Spectrom. Rev. \textbf{27}, 428 (2008).

\bibitem{Sun2015FP}B. H. Sun, Yu. A. Litvinov, I. Tanihata, Y. H. Zhang, Front. Phys. \textbf{10}, 102102 (2015).

\bibitem{Wang2017CPC}M. Wang, G. Audi, F. G. Kondev, W. J. Huang, S. Naimi, and X. Xu, Chin. Phys. C \textbf{41}, 030003 (2017).

\bibitem{Weizsacker1935ZP}C. F. Von Weizs\"{a}cker, Z. Phys. \textbf{96}, 431 (1935).

\bibitem{Bethe1937RMP}H. A. Bethe, R. F. Bacher, Rev. Mod. Phys. \textbf{8}, 82 (1936).

\bibitem{Moller2012PRL}P. M\"{o}ller, W. D. Myers, H. Sagawa, and S. Yoshida, Phys. Rev. Lett. \textbf{108}, 052501 (2012).

\bibitem{Wang2014PLB}N. Wang, M. Liu, X. Z. Wu, J. Meng, Phys. Lett. B \textbf{734}, 215 (2014).

\bibitem{Goriely2009PRLSkyrme}S. Goriely, N. Chamel, J.M. Pearson, Phys. Rev. Lett. \textbf{102}, 152503 (2009).

\bibitem{Goriely2016PRC}S. Goriely, N. Chamel, J. M. Pearson, Phys. Rev. C \textbf{93}, 034337 (2016).

\bibitem{Goriely2009PRLGogny}S. Goriely, S. Hilaire, M. Girod, S. P\'{e}ru, Phys. Rev. Lett. \textbf{102}, 242501 (2009).

\bibitem{Meng2006PPNP}J. Meng, H. Toki, S. G. Zhou, S. Q. Zhang, W. H. Long, and L. S. Geng, Prog. Part. Nucl. Phys. \textbf{57}, 470 (2006).

\bibitem{Vretenar2005PRp}D. Vretenar, A. V. Afanasjev, G.A. Lalazissis, P. Ring, Phys. Rep. \textbf{409}, 101 (2005).

\bibitem{Meng2016Book}International Review of Nuclear Physics, Vol. 10, Relativistic Density Functional for Nuclear Structure, edited by J. Meng (World Scientific, Singapore, 2016).

\bibitem{Meng2006PRC}J. Meng, J. Peng, S. Q. Zhang, and S. G. Zhou, Phys. Rev. C \textbf{73}, 037303 (2006).

\bibitem{Liang2008PRL}H. Z. Liang, N. Van Giai, J. Meng, Phys. Rev. Lett. \textbf{101}, 122502 (2008).

\bibitem{Niu2013PRCR}Z. M. Niu, Y. F. Niu, Q. Liu, H. Z. Liang, and J. Y. Guo, Phys. Rev. C \textbf{87}, 051303(R) (2013).

\bibitem{Niu2017PRC}Z. M. Niu, Y. F. Niu, H. Z. Liang, W. H. Long, and J. Meng, Phys. Rev. C \textbf{95}, 044301 (2017).

\bibitem{Sun2008PRC}B. Sun, F. Montes, L. S. Geng, H. Geissel, Yu. A. Litvinov, and J. Meng, Phys. Rev. C \textbf{78}, 025806 (2008).

\bibitem{Niu2009PRC}Z. M. Niu, B. H. Sun, and J. Meng, Phys. Rev. C \textbf{80}, 065806 (2009).

\bibitem{Niu2013PLB}Z. M. Niu, Y. F. Niu, H. Z. Liang, W. H. Long, T. Nik\v{s}i\'{c}, D. Vretenar, and J. Meng, Phys. Lett. B \textbf{723}, 172 (2013).

\bibitem{Xu2013PRC}X. D. Xu, B. Sun, Z. M. Niu, Z. Li, Y. Z. Qian, and J. Meng, Phys. Rev. C \textbf{87}, 015805 (2013).

\bibitem{Geng2005PTP}L. S. Geng, H. Toki, and J. Meng, Prog. Theor. Phys. \textbf{113}, 785 (2005).

\bibitem{Hua2012SCPMA}X. M. Hua, T. H. Heng, Z. M. Niu, B. H. Sun, J. Y. Guo, Sci. China Phys. Mech. Astron. \textbf{55}, 2414 (2012).

\bibitem{Arteaga2016EPJA}D. Pe\~{n}a-Arteaga, S. Goriely, and N. Chamel, Eur. Phys. J. A \textbf{52}, 320 (2016).

\bibitem{Kirson2008NPA}M. W. Kirson, Nucl. Phys. A \textbf{798}, 29 (2008).

\bibitem{Wang2011PRC}N. Wang and M. Liu, Phys. Rev. C \textbf{84}, 051303(R) (2011).

\bibitem{Niu2013PRCb}Z. M. Niu, Z. L. Zhu, Y. F. Niu, B. H. Sun, T. H. Heng, and J. Y. Guo, Phys. Rev. C \textbf{88}, 024325 (2013).

\bibitem{Zheng2014PRC}J. S. Zheng, N. Y. Wang, Z. Y. Wang, Z. M. Niu, Y. F. Niu, and B. Sun, Phys. Rev. C \textbf{90}, 014303 (2014).

\bibitem{Niu2016PRC}Z. M. Niu, B. H. Sun, H. Z. Liang, Y. F. Niu, and J. Y. Guo, Phys. Rev. C \textbf{94}, 054315 (2016).

\bibitem{Morales2010PRC}Irving O. Morales, P. Van Isacker, V. Vel\'{a}zquez, J. Barea, J. Mendoza-Temis, J. C. L\'{o}pez Vieyra, J. G. Hirsch, and A. Frank, Phys. Rev. C \textbf{81}, 024304 (2010).

\bibitem{Haykin2009Book}S. Haykin, Neural Networks and Learning Machines (Pearson Education, New Jersey, 2009).

\bibitem{Bishop2006Book}C. M. Bishop, Pattern Recognition and Machine Learning (Springer, Singapore, 2006).

\bibitem{Gazula1992NPA}S. Gazula, J. W. Clark, and H. Bohr, Nucl. Phys. A \textbf{540}, 1 (1992).

\bibitem{Gernoth1993PLB}K. A. Gernoth, J. W. Clark, and J. S. Prater, Phys. Lett. B \textbf{300}, 1 (1993).

\bibitem{Athanassopoulos2004NPA}S. Athanassopoulos, E. Mavrommatis, K. A. Gernoth, and J. W. Clark, Nucl. Phys. A \textbf{743}, 222 (2004).

\bibitem{Zhang2017JPG}H. F. Zhang, L. H. Wang, J. P. Yin, P. H. Chen, and H. F. Zhang, J. Phys. G: Nucl. Part. Phys. \textbf{44}, 045110 (2017).

\bibitem{Costiris2009PRC}N. J. Costiris, E. Mavrommatis, K. A. Gernoth, and J. W. Clark, Phys. Rev. C \textbf{80}, 044332 (2009).

\bibitem{Neal1996Book}R. Neal, Bayesian Learning of Neural Network (Springer, New York, 1996).

\bibitem{Utama2016PRC}R. Utama, J. Piekarewicz, and H. B. Prosper, Phys. Rev. C \textbf{93}, 014311 (2016).

\bibitem{Utama2016JPG}R. Utama, W. C. Chen, and J. Piekarewicz, J. Phys. G: Nucl. Part. Phys. \textbf{43}, 114002 (2016).

\bibitem{Kortelainen2010PRC}M. Kortelainen, T. Lesinski, J. Mor\'{e}, W. Nazarewicz, J. Sarich, N. Schunck, M. V. Stoitsov, and S. Wild, Phys. Rev. C \textbf{82}, 024313 (2010).

\bibitem{Goriely2016EPJA}S. Goriely, S. Hilaire, M. Girod, and S. P\'{e}ru, Eur. Phys. J. A \textbf{52}, 202 (2016).
\end{thebibliography}
\end{document}